\documentclass[a4paper]{jpconf}
\usepackage{graphicx}
\usepackage{subfig}
\begin{document}
\title{Comparison of entangling protocols in ABC-type spin chains}

\author{M. P. Estarellas$^1$, T. P. Spiller$^2$, I. D'Amico$^2$}

\address{$^1$ National Institute of Informatics, 2-1-2 Hitotsubashi, Chiyoda-ku, Tokyo 101-8430, Japan }
\address{$^2$ Department of Physics, University of York, YO10 5DD
York, United Kingdom}

%\ead{williams@ucl.ac.uk}

\begin{abstract}
In this contribution we consider an advantageous building block with potential for various quantum applications: a device based on coupled spins capable of generating and sharing out an entangled pair of qubits. Our model device is a dimerised spin chain with three weakly coupled embedded sites (defects). Three different entangling protocols were proposed for this chain in \cite{Estarellas:2017} and \cite{Wilkinson:2018}, one producing a Cluster state building block and two generating a Bell state, depending on the initial state injection. Here we compare the robustness of such protocols as quantum entangling gates against different types of fabrication (static energy fluctuations) and operation (timing injection delays) errors.
\end{abstract}

\section{Introduction}

Paul Busch was a real aficionado of the foundations and fundamentals of quantum theory. We all miss his deep thinking and wise words. Over recent decades, the work of Paul and many other quantum gurus has not only broadened our understanding and perspectives of quantum theory, but it has also opened up new potential for quantum applications in the real world. New quantum technologies are emerging, where the most fundamental features of quantum physics play centre stage, and provide advantage over current conventional technologies \cite{NandCh,UKNQTP}. Our contribution to this collection is about one such application.

Entanglement is perhaps the most fundamental ``non-classical" feature of quantum physics. Entanglement underpins numerous new quantum applications. It is the resource for teleportation \cite{Bennett1993}; it powers measurement-based quantum processing and computing \cite{Raussendorf2001,Walther2005}; it can be used for secure communications \cite{Ekert1991}; and it can be employed to sense and measure things better than we can do with conventional instruments \cite{Giovannetti2006,Dowling2008}. Generation and distribution of entanglement is therefore a very useful primitive, that can facilitate a whole range of quantum applications.

In this work we consider a model device, based on coupled spin-1/2 systems---or qubits, the building blocks of many new quantum technologies. One-dimensional systems of coupled spins, usually called chains, form versatile quantum devices. Spin chains can be employed to efficiently transport quantum information \cite{Bose2003,Bose2007}, even perfectly if the couplings between the spins can be tuned \cite{Christandl2004,Christandl2005,Plenio2004,Kay2010}, and such systems can be used to create, distribute and store entanglement \cite{Spiller2007,DAmico2007a}.

A very appealing reason for studying spin chains theoretically is that these investigations can apply to a range of physical systems, because qubits and their coupling mechanisms take many different forms. In the laboratory, the spins in a spin chain could be actual spins in a chain molecule, or nanoscale magnetic particles \cite{Tejada2001}, or spins in a string of coupled fullerenes \cite{Twamley2003}. Or they could be atoms held in optical lattices \cite{Atala2013,Meier2016}, ions in an ion trap \cite{Kielpinski2002,Monroe2014}, or electrons or excitons confined in coupled nanostructures \cite{DAmico2007,DAmico2006,Nikolopoulos2004}. Optical realisations, through coupled micro-pillars \cite{Jacqmin2014} or waveguides \cite{Blanco-Redondo2016}, and graphene-based systems \cite{Delplace2011} are also relevant. Thus spin chain modelling has applicability across a broad spectrum of qubit systems, and thus potential future technology platforms.

\section{ABC-type spin chains and entanglement generation protocols}

\begin{figure}[ht!]
\centering
  \includegraphics[width=0.6\textwidth]{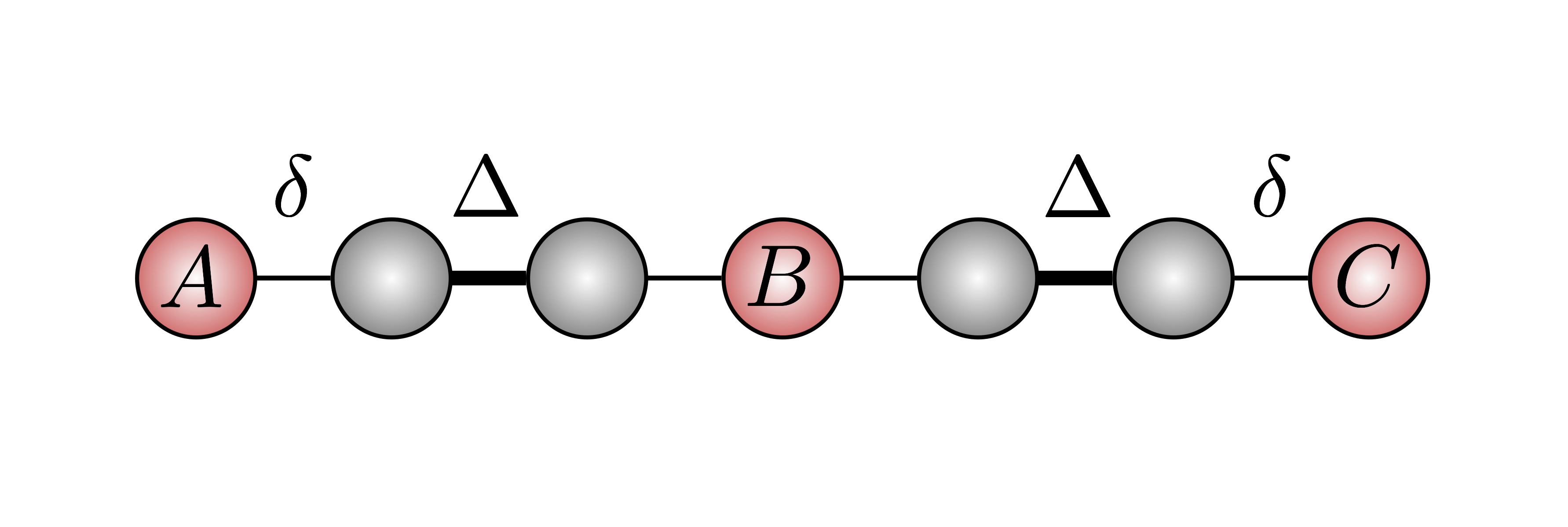}
\caption{ABC-type spin chain configuration.}
\label{ABC}
\end{figure}

The model system we consider for our entanglement generation protocols is a spin chain with alternating weak ($\delta$) and strong ($\Delta$) couplings, distributed such that there are three -defect- sites (labelled $A$, $B$ and $C$) weakly coupled to the rest of the chain as shown in Fig. \ref{ABC}. For protocol demonstration, here we focus on a spin chain configuration with $N=7$ sites. However, longer chains could be considered for implementation of the same protocols \cite{Estarellas:2017,Wilkinson:2018}.

This spin chain can be described by the following time-independent Hamiltonian

\begin{eqnarray}
\label{hami}
{\cal{H}} = \sum_{i=1}^{N}\epsilon_{i}|1\rangle \langle 1|_{i} + \sum_{i=1}^{N-1} J_{i,i+1}[ |1\rangle \langle 0|_{i} \otimes |0\rangle \langle 1|_{i+1} + h.c.],
\end{eqnarray}
with the coupling $J_{i,i+1}$ equal to either $\Delta$ or $\delta$ depending on the site (see Fig.\ \ref{ABC}) and the on-site energy  $\epsilon_{i}=0$ unless stated otherwise (i.e. when diagonal random disorder is added). For each spin qubit, the standard computational basis states are represented by spin up, $|1\rangle$, and spin down, $|0\rangle$, with the zero (excitation) state of the system being all spins down. We note that this zero state could be prepared as the ground state of the system (brought into contact with a dissipative environment at temperature $T$) in a suitable external magnetic field that makes all the on-site energies sufficiently positive, so $\epsilon_{i} \gg kT$ for all $i$.  Or potentially through projective measurement of every spin, with local spin flips applied to each that is measured up, although this second method would clearly require local addressing and measurements at each site. We note also that the Hamiltonian (\ref{hami}) preserves excitation (spin up) number, so if amplitudes of different excitation (spin up) numbers are prepared from the zero state, these then evolve separately according to the dynamics for that excitation number sector.

It has been demonstrated \cite{estarellas2016,huo2008,Almeida2016,Gualdi2011} that related dimerised chains have high fidelity quantum state transfer (QST) properties. In general QST properties in spin chains can be exploited to generate entanglement, for example as discussed in \cite{Yung2005,Clark2005,Clark2007,Alkurtass2014,Banchi2015}. This is something that we exploit in our protocols  in order to generate the desired entangled states in dimerised chains \cite{Estarellas:2017, Wilkinson:2018}. The operation of these spin chain devices relies on their natural dynamics. A simple initial state of the system is prepared and the protocol is then effected by evolution under ${\cal{H}}$ for some pre-determined time.

\begin{figure}[h!]
\centering
  \includegraphics[width=0.7\textwidth]{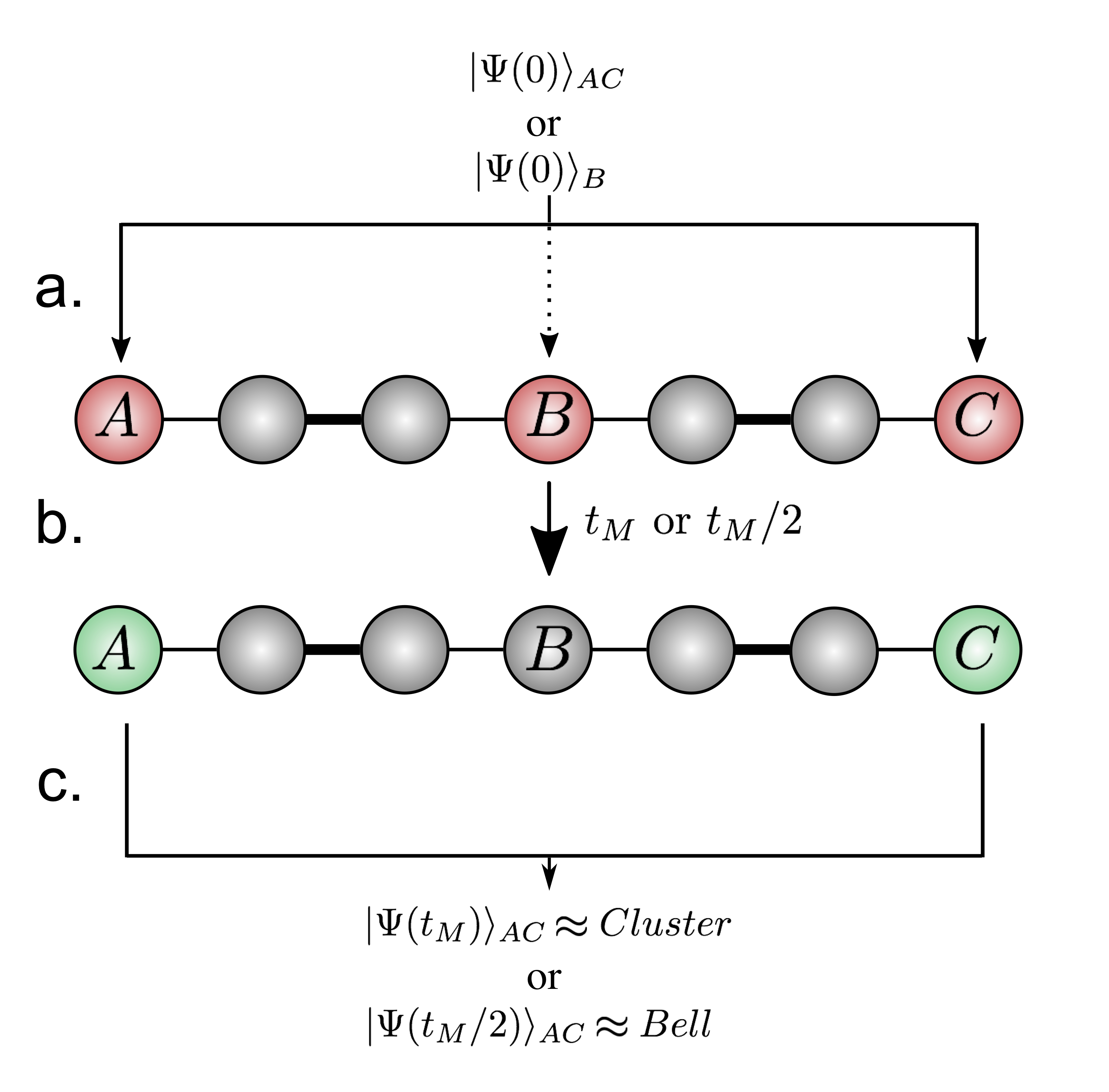}
\caption{Entangling protocols: (a) Initial injection (red) at sites A and C (solid arrows) or at site B (dashed arrow); (b) Evolution of the system up to time $t_E\approx t_M$ or $t_E\approx t_M/2$; (c) Generation of a highly entangled Cluster state or Bell state, respectively, between sites A and C (green).}
\label{protocol1}
\end{figure}

\begin{figure}[]
	
	\begin{center}
		\includegraphics[width=0.45\textwidth]{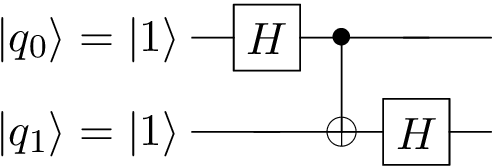}
		\caption{\label{ClusterCircuit}Cluster state creation written as a circuit. $H$ is the single-qubit Hadamard gate and the two-qubit gate is a CNOT \cite{NandCh}.}%
		%\label{ClusterCircuit}%
	\end{center}
	%\end{figure}
	%\begin{figure}%
	\begin{center}
		\includegraphics[width=0.39\textwidth]{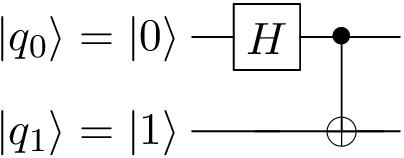}
		\caption{\label{bellCircuit}Bell state creation written as a circuit. $H$ is the single-qubit Hadamard gate and the two-qubit gate is a CNOT \cite{NandCh}.}%
		%\label{bellCircuit}%
	\end{center}
\end{figure}

The entanglement generation protocols are schematically illustrated in Fig.\ \ref{protocol1}.  The initial injection, applied to the zero state of the system, defines the overall initial state of the chain, and hence the final entangled state. We note that injection is one of only two interactions the user has to make with the system. Initialization is done by either simultaneously injecting a double excitation state at sites $A$ and $C$, or by a single excitation state injection at site $B$. Depending on the chosen initial state, the entangled state generated will either be approximated to a Cluster state building block (a superposition across the zero-, one- and two-excitation sectors) or a Bell state. The second user interaction with the system involves either retrieving the entangled state at a specific pre-determined time later in the dynamics, or performing a measurement. The Cluster state building block is so named because repeated application of such a protocol can be used to knit larger cluster states \cite{ronke2011_2}. From now on, with this understood, we simply refer to this two-qubit entanglement example as a Cluster state.

Let us now look in more detail at the injection for the generation of this two-qubit Cluster state, denoted approach (i). In the ideal case, this is a maximally entangled state formed by an equal superposition of all the involved site basis vectors. Fig.\ \ref{ClusterCircuit} shows the circuit equivalent to an ideal Cluster state-generation gate.  In this case, the protocol is initiated at $t=0$ with the injection of two initial $|+\rangle=\frac{1}{\sqrt{2}}(|0\rangle+|1\rangle)$ states at the chain ends (sites $A$ and $C$).
We can write the initial state in the standard basis as follows:

%\begin{eqnarray}	
%|\Psi(0)\rangle_i &= &\frac{1}{2}\Big(|+\rangle_A\otimes|+\rangle_C\Big)\otimes|0\rangle_{rest-of-chain}=\frac{1}{2}\Big(|0\rangle_A|0\rangle_C+|1\rangle_A|0\rangle_C+|0\rangle_A|1\rangle_C+ \nonumber\\
%& & |1\rangle_A|1\rangle_C\Big)\otimes|0\rangle_{rest-of-chain}.
%\label{initial}
%\end{eqnarray}

\begin{eqnarray}	
\textrm{i) }|\Psi(0)\rangle_i &= &\frac{1}{2}\Big(|+\rangle_A\otimes|+\rangle_C\Big)\otimes|0\rangle_{rest-of-chain} \nonumber\\ &= &\frac{1}{2}\Big(|0\rangle_A|0\rangle_C+|1\rangle_A|0\rangle_C+|0\rangle_A|1\rangle_C+
|1\rangle_A|1\rangle_C\Big)\otimes|0\rangle_{rest-of-chain}.
\label{initial}
\end{eqnarray}

We turn now to the injection protocols for Bell state creation. The  circuit equivalent to the Bell state-generation gate is presented in Fig.\ \ref{bellCircuit}. There are two approaches: (ii) the simultaneous injection of two excitations at sites $A$ and $C$; or (iii) single-excitation injection at site $B$. These possibilities present different features and thus distinct advantages. The single injection at the centre, (iii), allows generation and distribution of a Bell state with the convenience and ease of having to initially interact with one site only (site $B$). Alternatively, some applications such as modular quantum processor proposals may require generation of this same state with simultaneous compliance and contribution of two separate parties or quantum registers \cite{Gualdi2011}. Then the appropriate injection approach is (ii), which involves the two parties $A$ and $C$, who will need to agree a priori initiation of the protocol. For Bell state creation, we can write the state of the chain at initial time, $t=0$, as follows:

%\begin{equation}
%\textrm{ii) }
%|\Psi(0)\rangle_{ii}=\big(|1\rangle_A\otimes|1\rangle_C\big)\otimes|%0\rangle_{rest-of-chain}.
%\end{equation}
%\begin{equation}
%\textrm{iii) } |\Psi(0)\rangle_{iii}=|1\rangle_B\otimes|0\rangle_{rest-of-%chain}.
%\end{equation}

\begin{eqnarray}
\textrm{ii) }\;
|\Psi(0)\rangle_{ii}\; &= &\big(|1\rangle_A\otimes|1\rangle_C\big)\otimes|0\rangle_{rest-of-chain}, \\
\textrm{iii) } |\Psi(0)\rangle_{iii} &= &|1\rangle_B\otimes|0\rangle_{rest-of-chain}.
\end{eqnarray}

After initialization, the system is allowed to evolve under its natural dynamics to a given time,  the `entangling time', $t_E$, which depends on the version of the protocol used and the required state to be generated (Cluster or Bell state). For the Cluster state generation \cite{Estarellas:2017}, $t_E$ is the mirroring time ($t_E\approx t_{M}$), i.e. the time needed for an arbitrary initial state to evolve to its mirror position with respect to the centre of the chain. The reason an initial product state `mirrors' to an entangled state is because of an additional fundamental sign change that arises for (just) the two-excitation amplitude \cite{Yung2005,Clark2005,Clark2007}. This sign change, which effectively arises from the exchange of two fermions, underpins the generation of a Cluster state building block \cite{ronke2011_2}.

For the Bell state generation \cite{Wilkinson:2018}, the `entangling time' is approximately half the mirroring time ($t_E\approx t_M/2$). In case (iii), at this time a single excitation injected at the centre effectively delocalises to a superposition of being at both ends of the chain \cite{Wilkinson:2018,Banchi2015}. In case (ii), at this time the state of the chain becomes a product state of such a delocalised single excitation with a state of the rest of the chain containing the other excitation \cite{Wilkinson:2018,Alkurtass2014}.

In all cases, at $t_E$ the system state becomes highly entangled between sites $A$ and $C$, from which the two entangled qubits can be extracted, if desired \footnote{An alternative to this is to effectively freeze and then protectively store the state using a slightly modified system configuration and the protocol, see \cite{Estarellas:2017}.}. In Table \ref{statestab} we summarize the three different initial states we employ and their corresponding injection sites, as well as the generated entangled states and the associated entangling times, $t_E$.

\begin{table}[h]
\caption{\label{statestab} Initialization states (injection), their corresponding entangled states, and their corresponding entangling times $t_E$. The $|+\rangle$ state corresponds to $\frac{1}{\sqrt{2}}(|0\rangle+|1\rangle)$. For simplicity, only the states of the relevant sites ($A$, $C$ or $B$) are presented and a compact notation is used.}
\begin{center}
\begin{tabular}{c|c|c|c}
\br
& \bf{Initial state} & \bf{Entangled State} & $\mathbf{t_E}$ \\\hline
\bf{i} & \small $|\Psi(0)\rangle_{AC}=(|+\rangle_A\otimes|+\rangle_C)$  & \small $|\Psi(t_E)\rangle_{AC}\approx\frac{1}{2}(|00\rangle_{AC}-|01\rangle_{AC}-|10\rangle_{AC}-|11\rangle_{AC})$ & $t_M$ \\\hline
\bf{ii} & \small $|\Psi(0)\rangle_{AC}=|1\rangle_A\otimes|1\rangle_C$ & $|\Psi(t_E)\rangle_{AC}\approx\frac{-i}{\sqrt{2}}(|10\rangle_{AC}+|01\rangle_{AC})$ & $t_{M}/2$ \\\hline
\bf{iii} & \small $|\Psi(0)\rangle_B=|1\rangle_B$ & $|\Psi(t_E)\rangle_{AC}\approx\frac{-i}{\sqrt{2}}(|10\rangle_{AC}+|01\rangle_{AC})$ & $t_{M}/2$\\
\br
\end{tabular}
\end{center}
\end{table}

\section{Comparison of the entangling protocols}
\begin{figure*}[ht!]
\centering
%\resizebox{0.8\textwidth}{!}{
  \includegraphics[width=0.8\textwidth]{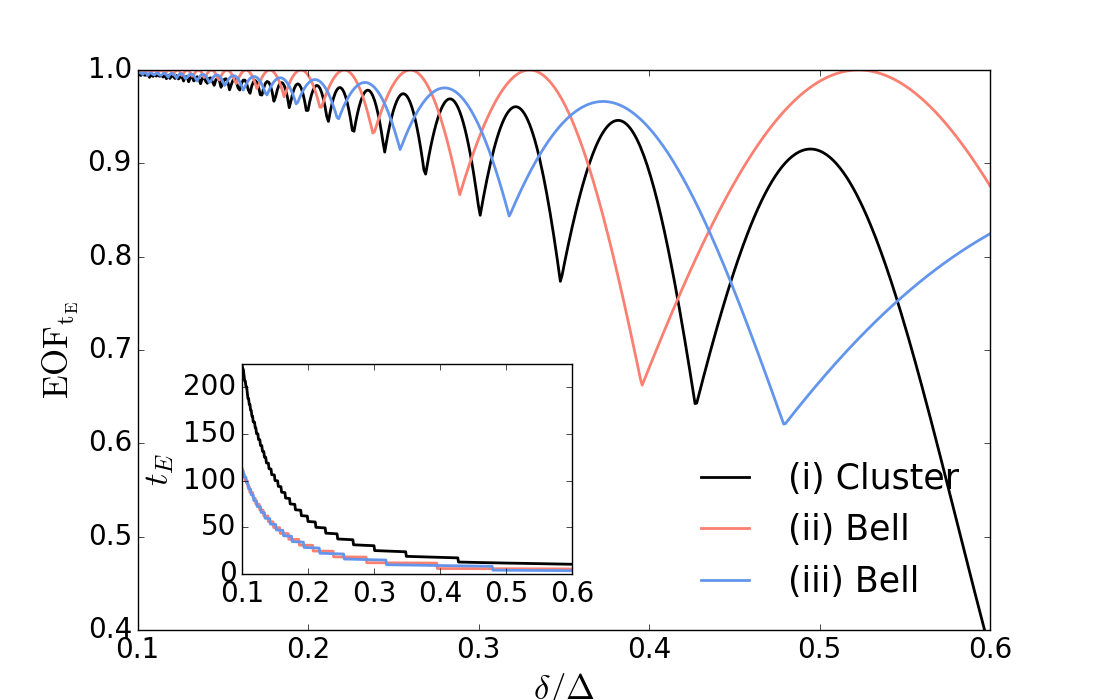}
%}
\caption{Entanglement of formation at $t_{E}$ for the three different protocols, (i), (ii) and (iii), against varying coupling ratio, $\delta/\Delta$, for an $ABC$ chain with $N=7$. The inset shows how the entangling times, $t_E$, vary with the coupling ratio.}
\label{eofVSratio}
\end{figure*}

The chain mirroring time $t_M$, and hence the entangling time $t_E$, can be estimated analytically in terms of the coupling ratio $\delta/\Delta$, and it decreases as this ratio increases \cite{Estarellas:2017, Wilkinson:2018}. Intuitively this can be understood by considering the smaller $\delta$-couplings to form `bottle-necks' for the excitation dynamics, with respect to those of a uniform chain with all $\Delta$-couplings. Of course, due to quantum interference, we cannot think of the excitation dynamics as a simple flow between contiguous sites, so increasing  $\delta/\Delta$ also affects the level of entanglement and in a non-monotonic way. Fig.\ \ref{eofVSratio} presents the behavior of the three protocols for different coupling ratios. Here the exact entangling times, $t_E$, for each ratio are calculated numerically by determining when the entanglement peaks at a maximum for that ratio value. The entanglement is measured as entanglement of formation ($EOF$) \cite{Wooters:1998} of the reduced state for the sites of interest in the chain, with $EOF\in[0,1]$ and $EOF=1$ indicating a maximally entangled state. This $EOF$ therefore implicitly includes any damage due to the sites of interest having residual entanglement with other sites. The exact entangling times $t_E$ are still close to $t_M$ for the Cluster state and $t_M/2$ for the Bell state protocols. As would be expected, with an increasing value of the coupling ratio, $\delta/\Delta$, the entangling protocols are indeed faster, as observed in the inset of Fig.\ \ref{eofVSratio}. However the corresponding value for the entanglement has an oscillatory behaviour, with well-defined peaks whose maxima decrease with increasing  coupling ratio $\delta/\Delta$ for two protocols out of three. The (ii) protocol, corresponding to creation of Bell states from two initial injections, indeed continues to achieve (almost) maximal entanglement at every peak.

Ideally, one would desire maximal entanglement at a short $t_E$ (to optimise entanglement delivery in the presence of decoherence processes occuring in the device), as well as some level of robustness against static fabrication defects and/or injection errors. In the rest of this section, we will seek a compromise between the entangling times, $t_E$, the $EOF$ of the state obtained at those times, and the robustness of the protocol.

Having already analysed the schemes and presented results for the ideal cases, we next turn to study the behaviour of the protocols against fabrication or injection errors with increasing coupling ratio. For these analyses we consider the overall protocols for a selection of three different coupling ratios, corresponding to different maxima of each profile of Fig.\ \ref{eofVSratio}, as summarized in Table\ \ref{ratios}.

\begin{table}[h]
\caption{\label{ratios} Selected coupling ratios $\delta/\Delta$ corresponding to some chosen maxima of the Fig.\ \ref{eofVSratio} profiles for protocols (i), (ii) and (iii).}
\begin{center}
\begin{tabular}{lll}
%\begin{tcolorbox}[tab2,tabularx*={\renewcommand{\arraystretch}{1.5}}{X|X| X},title={Coupling ratios},boxrule=0.9pt]
\br
\bf{i. Cluster state} & \bf{ii. Bell state} & \bf{iii. Bell state}\\
\mr
0.205 & 0.260 & 0.204 \\\hline
0.382 & 0.330 & 0.280 \\\hline
0.490 & 0.523 & 0.373 \\
%\end{tcolorbox}
\br
\end{tabular}
\end{center}
\end{table}

We will test the protocols' robustness against static random perturbations of both on-site energies $\epsilon_i$ (diagonal disorder) and couplings $J_{i,i+1}$ (off-diagonal disorder) in the Hamiltonian (\ref{hami}). In addition, we will consider the effects of relative time delays between the injections of the two excitations in protocols (i) and (ii). Details of the methods and corresponding results are described below.

\subsection{Diagonal and off-diagonal disorder}
\label{sec:diag}

Local energy fluctuations may be induced by fabrication defects and/or local field fluctuations, linked to the specific physical realisation of the spin chain. We simulate static (compared to the timescale of the entanglement dynamics) energy fluctuations by adding random disorder to the diagonal terms of the Hamiltonian. Diagonal disorder has been shown to be one of the most damaging sources of static decoherence for the single excitation sector in spin chains \cite{ronke2011_1}.

The first term on the right hand side of the Hamiltonian (\ref{hami}) is set to have $\epsilon_{i}=Er_{i,i}\delta$, where $r_{i,i}$ is a random number from a uniform distribution between $-1/2$ and $1/2$, $E$ is a dimensionless positive parameter that fixes the scale of the disorder, and $\delta$ is the weak coupling.

Our second approach to model fabrication errors and local defects is accounting for static (again, compared to the timescale of the entanglement dynamics) errors in the off-diagonal terms of the Hamiltonian (\ref{hami}). Such perturbations represent random disorder in the couplings of the chain, $J_{i,i+1}$. In order to do so we modify the second term on the right hand side of the Hamiltonian  (\ref{hami}) by setting $J_{i,i+1}^{eff}=J_{i,i+1}+Er_{i,i+1}\delta$, where $r_{i,i+1}$ are again randomly generated from a uniform distribution.

For each type of disorder, each protocol, and each selected coupling ratio, we calculate 1000 random realisations and then average over the values of the $EOF$ to estimate the robustness of the protocols.
\begin{figure}%
    \begin{center}
%    \resizebox{\textwidth}{!}{
    \subfloat{{\includegraphics[width=0.8\textwidth]{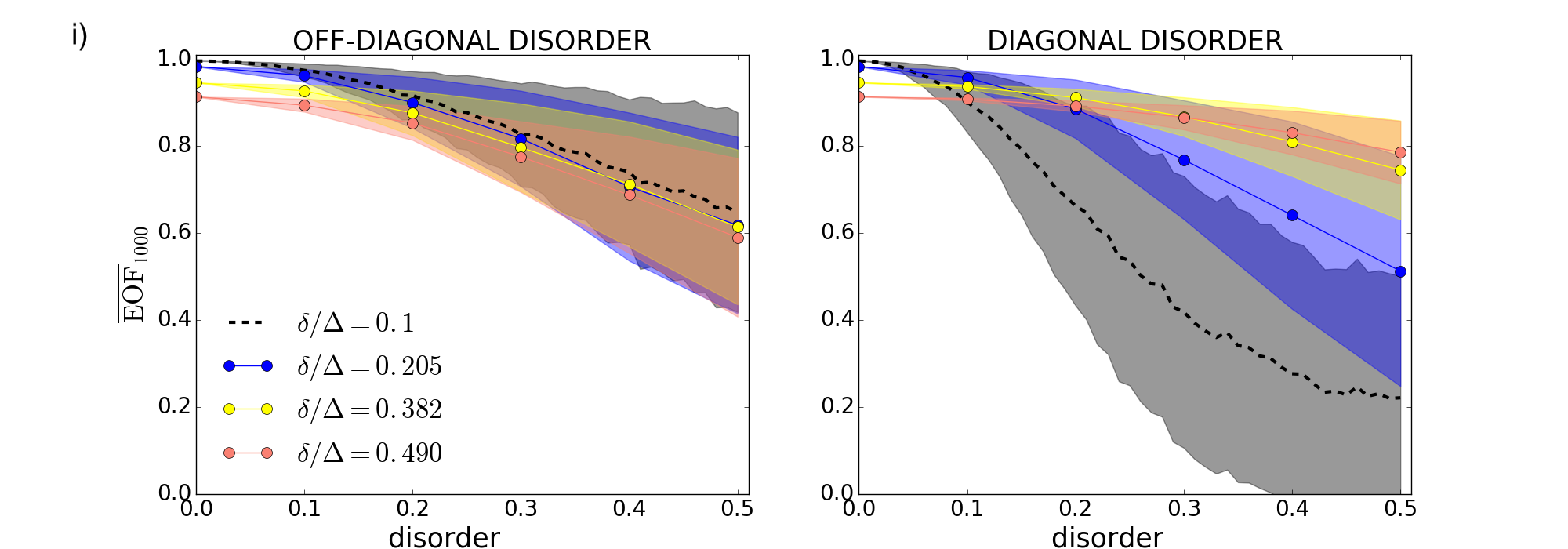} }}
%    }
\end{center}
    \begin{center}
%    \resizebox{\textwidth}{!}{
    \subfloat{{\includegraphics[width=0.8\textwidth]{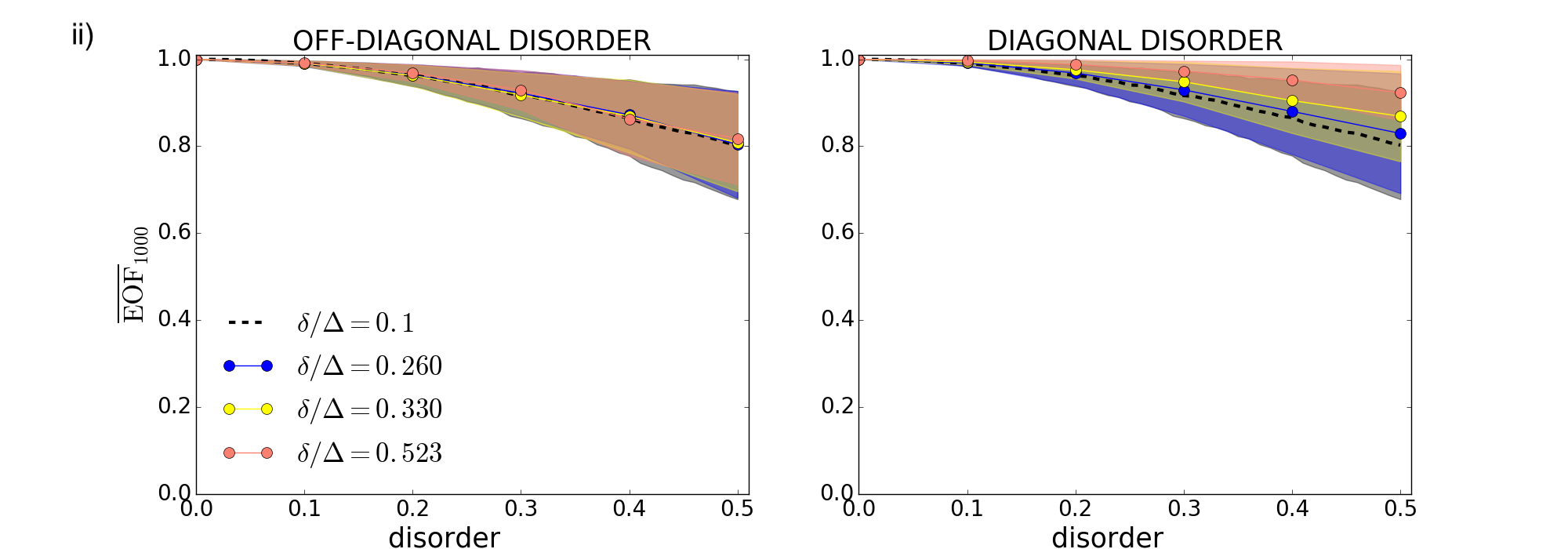} }}
%    }
\end{center}
    \begin{center}
%    \resizebox{\textwidth}{!}{
    \subfloat{{\includegraphics[width=0.8\textwidth]{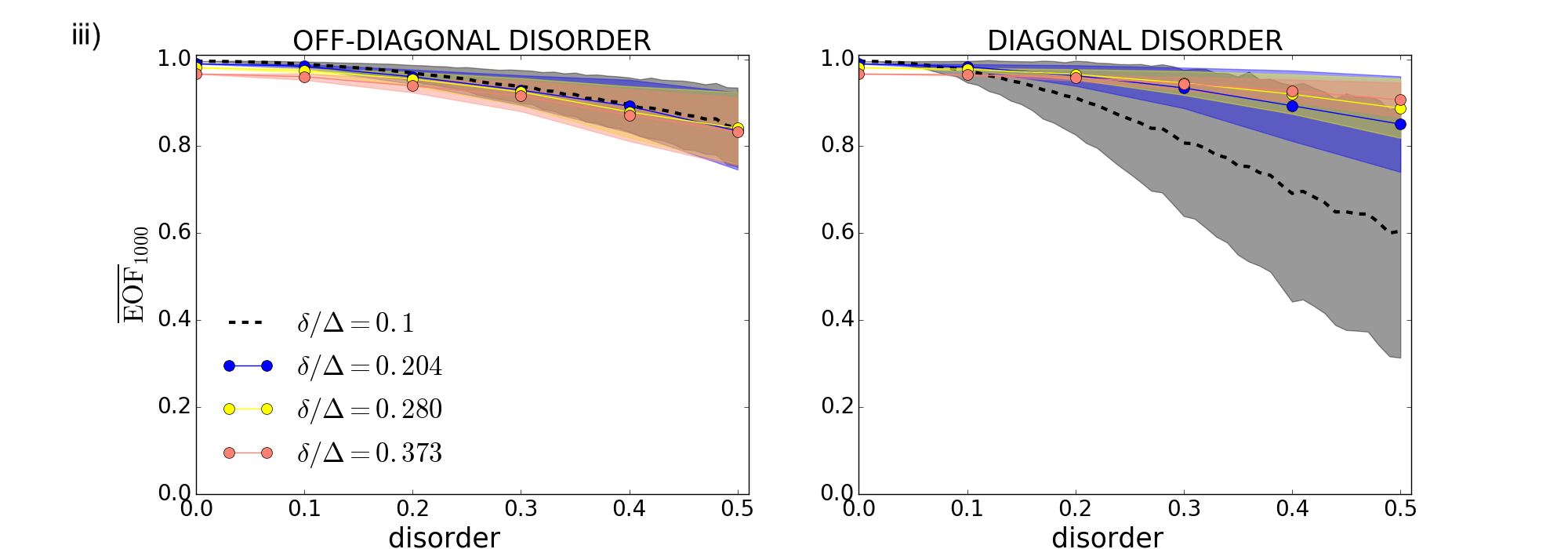} }}%
%    }
\end{center}
    \caption{Averaged $EOF$ at the unperturbed $t_E$ for different levels of off-diagonal (l.h.s) and diagonal (r.h.s) disorders and coupling ratios from Table \ref{ratios} for each of the three protocols (i), (ii) and (iii). Averages are calculated over 1000 random realizations. The black dashed line shows the averaged $EOF$ for a ratio $\delta/\Delta=0.1$ and orange, yellow and blue solid lines show the other three compared ratios, as labelled. The shadows represent the standard deviation of a single realization for each averaged profile.}%
    \label{eofVSnoise}%
\end{figure}
The various panels of Fig.\ \ref{eofVSnoise} show the averaged $EOF$  obtained at the unperturbed $t_E$, for each of the studied coupling ratios and against an increasing level of diagonal and off-diagonal disorder (up to 50\% of the weak coupling), for protocols (i), (ii) and (iii). We observe that even for coupling ratios as high as $0.4\sim 0.5$ our protocols are still considerably resilient against static perturbations. Importantly, for diagonal disorder---previously thought to be the most damaging---the performance of all protocols is dramatically improved by an increasing coupling ratio. So {\it in all three cases} (i), (ii) and (iii), for $\delta/\Delta>0.37$ the averaged $EOF$ at $t_E$ is always close or even greater than 0.8, even with a diagonal disorder at 50\% of the weak coupling value. Additionally, for both types of disorder and all protocols, as the coupling ratio is increased, each realisation for the $EOF$ calculation deviates less from the averaged value, as shown by the narrower standard deviation shades from Fig.\ \ref{eofVSnoise}. All of these analyses offer clear evidence that our protocols can be optimized, in terms of making the entangling operation times faster without necessarily sacrificing the quality of the entangled state. Hence it is possible to offer optimal and useful $EOF$ values but also with resilience against actual system disorder.

\subsection{Time Delays}

For protocols (i) and (ii), the ideal scenario is synchronous injection at sites $A$ and $C$.  We want to assess how damaging it is for the target entangled state when the excitations are injected in an asynchronous manner. This is modelled by adding a time delay $t_D=D t_E$ between the two injections, where $D$ is the dimensionless scale of the delay ($0\le D\le0.1$ in this work) and $t_E$ is the unperturbed entangling time, different for each protocol and coupling ratio. The justification for investigating delays at a small fraction of $t_E$ is that any viable physical implementation will always need to operate with accuracy at the timescale of its expected $t_E$. Therefore it is appropriate to investigate a window of injection (and extraction) timing errors covering a small fraction of the relevant $t_E$.

\begin{figure*}[ht!]
\centering
  \includegraphics[width=1\textwidth]{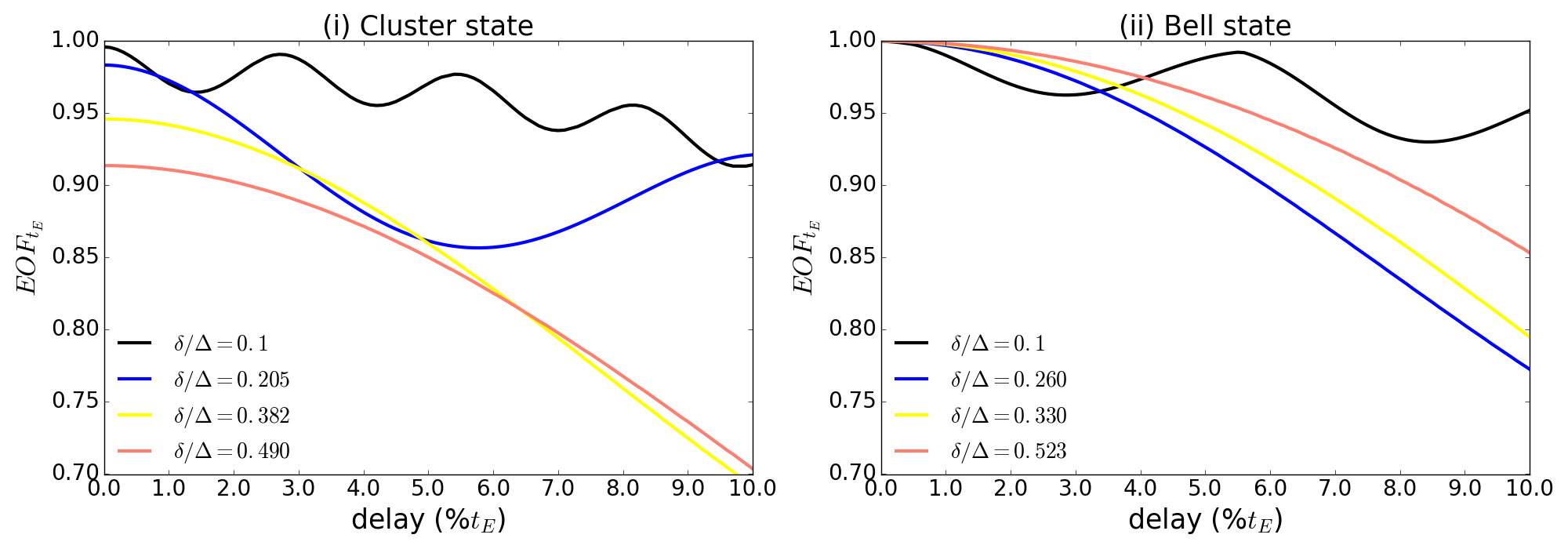}
\caption{$EOF$ at $t_E$ against the input delay as a
fraction of the entangling time ($t_E$) for protocols (i) (l.h.s) and (ii) (r.h.s) with asynchronous injections of $|+\rangle$ and $|1\rangle$ respectively at sites $A$ and $C$, and the different chosen coupling ratios, $\delta/\Delta$, depicted in Table \ref{ratios}. }
\label{delay}
\end{figure*}

The initial injection is then performed in two parts. First, we initiate the time evolution with injection at site $A$, so that the initial state becomes $|+,0...0,0\rangle$ for protocol (i) or $|1,0...0,0\rangle$ for protocol (ii). Immediately before the second, delayed, injection time, so at $t_D^-$, we retrieve the overall state of the system $|\Psi(t_D^-)\rangle$. We can write $|\Psi(t_D^-)\rangle=\sum_i^{N}c_{i}|\Phi_i^{1ex}\rangle + c_{0}|0,0...0,0\rangle$, with $N$ being the total number of sites, $c_0=1/\sqrt{2}$ for protocol (i) and $c_0=0$ for protocol (ii), and where $\sum_i^{0,N}|c_i|^2=1$ and $|\Phi_i^{1ex}\rangle=|0,0..1_i..0,0\rangle$. At time $t_D^+$ and for protocol (ii), injection at site $C$ is simulated by mapping the coefficients $c_i$ into the two-excitation vectors, $|\Phi_v^{2ex}\rangle=|00..1_i1_b..00\rangle$ with $i\neq b$, to find $|\Psi(t_D^+)\rangle_{(ii)}=\sum_vc_i|\Phi_v^{2ex}\rangle$.
For protocol (i), the coefficients $c_i$ must be mapped into the zero, one and two excitation sectors. For the zero-excitation sector, it is enough to renormalise the whole state with an additional $1/\sqrt{2}$ factor; for the single-excitation sector, one needs to account for the state with a single excitation at site $C$; and for the two-excitation sector we proceed as in protocol (ii), such that $|\Psi(t_D^+)\rangle_{(i)}=1/\sqrt{2}(|\Psi(t_D^-)\rangle+\vert 0,0...0,1_C\rangle+\sum_vc_i|\Phi_v^{2ex}\rangle)$.
We note that in our simulations the time delay is short enough so that at $t_D^-$ there is negligible probability for site $C$ to be occupied\footnote{This probability is at most $2\times 10^{-8}$ and $9\times 10^{-12}$, for the Cluster and Bell state generation respectively, corresponding to the scenarios with larger coupling ratios (red curves in Fig.\ \ref{delay}) and a delay of 10\%.}, hence the aforementioned treatments are an accurate approximation and no state renormalisation is required. In cases when the occupation probability of an ideally empty site is non-negligible at the time of the second, delayed injection, it is clearly important to consider the actual injection mechanism. The actual state prepared will then clearly depend on if the mechanism is some sort of SWAP operation at the relevant site, or a spin flip operation, or some kind of projection. A detailed discussion of injection mechanisms is given in \cite{ronke2011_1}.

 The effect of time delay errors on protocols (i) and (ii) is shown in Fig.\ \ref{delay} for the different chosen coupling ratios from Table \ref{ratios}. For $\delta/\Delta=0.1$, we observe that even with an injection time delay of 10\% $t_E$, the $EOF$ is still very high for both protocols, with values of 0.91 and 0.95, respectively. This robustness decreases with an increasing coupling ratio for both protocols. 
As $\delta/\Delta$ increases, the eigenstates peaking at sites $A$, $B$, and $C$ are less localized \cite{Wilkinson:2018}. This implies that when injecting at these sites, the excitation is not injected in a single eigenstate (as it would be for $\delta/\Delta\to 0$), but in
a superposition of eigenstates with increased weight of the additional
eigenstates as $\delta/\Delta$ increases. This introduces additional phases in the wavefunction, and hence a more complicated dynamics, so that a delayed second injection affects the desired dynamics more, and the performance worsen with increasing $\delta/\Delta$.
However, the $EOF$ still remains high, with values of 0.70, and 0.77 for the larger ratios ($\delta/\Delta=0.490$ and $\delta/\Delta=0.523$), respectively,  and a time delay of 10\% $t_E$.
 We can conclude that our protocol is therefore also considerably robust against asynchronous state injections, which one would expect to be a small fraction of the relevant $t_E$ for good physical implementations.

We finish this section by commenting on the scalability of the system. The chain length in this model can be increased by adding sets of two dimers, one either side of site $B$ to preserve the symmetry. The resulting expanded systems will still support all the protocols discussed here. However, the time taken for entanglement creation would increase exponentially with chain length \cite{Wilkinson:2018}. The important feature of the application discussed in this work is the robust creation of entanglement and its usefulness for modular quantum computation schemes. Scalability with chain length is less of an issue for this, in comparison to applications that utilise spin chains as quantum communication buses.

\section{Conclusions}

In this contribution we have compared the robustness of the entangling protocols proposed in \cite{Estarellas:2017} and \cite{Wilkinson:2018} with respect to fabrication errors, slowly fluctuating fields and asynchronous  injection of excitations during initialization. For all protocols and all forms of error considered, the resulting entanglement of formation has been shown to be very robust, even when the maximum percentage error considered is very large (e.g. up to 50\% of the relevant Hamiltonian parameters for fabrication and slowly fluctuating fields).  Energy fluctuation (diagonal) errors are generally more damaging than coupling (off-diagonal) disorder (against which there is excellent robustness) for small coupling ratios. So for practical implementations it is most important to focus on the reduction of fabrication errors that give rise to diagonal disorder. Timing injection errors, up to 10\% of the time needed for producing entanglement, reduce the entanglement of formation by only a few \%  for $\delta/\Delta=0.1$, with a more damaging effect when the coupling ratio is increased. However, good physical implementations would always be expected to have timing control errors limited to at most a small fraction of the actual device operation time.

We have provided a systematic comparison of the effect of imperfections on a large range of characteristic coupling ratios: this is important  due to the exponential speedup of all entanglement protocols as the coupling ratio increases \cite{Estarellas:2017,Wilkinson:2018}.
We find that for all protocols the robustness to off-diagonal disorder is almost independent of the coupling ratio; even more exciting, the robustness to diagonal disorder increases significantly as the coupling ratio is increased. This result demonstrates a positive feedback between the speed and robustness to disorder of the protocols. The ability to maximize both of these crucial factors at little to no reduction in the maximum $EOF$ allows us to propose $ABC$-type chains as rapid and reliable entanglement generation devices. $ABC$-type chains could therefore be of use in applications utilising various quantum computer architectures and across a variety of physical platforms, particularly where `off-line' and robust entanglement creation and distribution between two parties is required as a resource.

\end{document}